\documentclass[final,3p,times,twocolumn]{elsarticle}

\usepackage{graphics}
\usepackage{graphicx}
\usepackage{epsfig}
\usepackage{amssymb}
\usepackage{amsfonts}
\usepackage{amssymb}
\usepackage{amsmath}
\usepackage{lipsum}
\usepackage{graphicx}
\usepackage{fixfoot}
\usepackage{color}
\usepackage{amsthm}
\usepackage{lineno}
\usepackage{amsfonts}
\usepackage{amsthm}
\usepackage{amssymb}
\usepackage{amsmath}
\usepackage{graphicx}
\usepackage{natbib}
\usepackage{verbatim}
\usepackage{amsfonts}
\usepackage{amsthm}
\usepackage{amssymb}
\usepackage{amsmath}
\usepackage{graphicx}
\usepackage{natbib}
\usepackage{verbatim}
\usepackage{color}
\usepackage{bm}

\journal{Materials Letter}

\begin{document}

\begin{frontmatter}
\title{ZrO$_2$ tapes as an alternative low-cost flexible substrate to artificially nanostructured materials}
\author[ufrn1]{M. A. Correa}
\ead{marciocorrea@dfte.ufrn.br}
\author[ufrn2]{M.~R.~Araujo}
\author[ufrn1,ufrn2]{W.~Acchar}
\author[ufrn1]{A.~L.~R.~Souza}
\author[ufrn1]{A.~S.~Melo }
\author[ufrn1]{F.~Bohn}

\address[ufrn1]{Departamento de F\'{i}sica, Universidade Federal do Rio Grande do Norte, 59078-900 Natal, RN, Brazil}
\address[ufrn2]{Departamento de Engenharia de Materiais, Universidade Federal do Rio Grande do Norte, 59078-900 Natal, RN, Brazil}

\begin{abstract}
We associate tape casting and magnetron sputtering techniques to engineer flexible nanostructures using ZrO$_2$ green tape as substrate.
We systematically investigate the structural, magnetic and electrical properties of NiFe/Cr/NiFe trilayer nanostructures, with variable thickness values of the NiFe and Cr layers, grown onto rigid glass and flexible ZrO$_2$ tape substrates.
We verify the mirroring of these properties in the trilayer nanostructures, irrespective on the kind of employed substrate.
The fact that the trilayer nanostructures can be reproduced in distinct substrates corresponds to an important advance for their applicability. 
The results place the ZnO$_2$ green tape as an attractive candidate for application as low-cost flexible substrate in the development of electrical and magnetic sensor elements with high sensitivity to mechanic stress.
\end{abstract}


\end{frontmatter}



\section{Introduction}
\label{Introduction}

Low-cost thin ceramic membranes 
present interesting properties and find a wide range of applications in the fields of electronics and chemical processing technology.
The most widely used techniques for the fabrication of these membranes are hot pressing, hot rolling and tape casting. 
Thin ceramic materials have been developed intensively in the last decades by the tape casting method, which are a low cost process and particularly indicated for multilayered ceramic composite materials and solid oxide fuel cells (SOFC)~\cite{AAC114p127,JACS93p1313,CI41p7836}. 
Tape casting is a widespread colloidal processing that has the advantage to produce homogeneous green structures. 
This process consists basically in forming slurry and in casting it through a doctor-blade on a generally moving surface~\cite{MSEA202p206,RPJ17p424}. 
The suspension is comprised by a dispersion of a ceramic powder in a solvent, with the addition of binders, plasticizers and dispersants~\cite{JECS19p1725,JACS83p1557,JECS30p3397,IJACT12p18}. 
In the last years aqueous-based tape casting has been investigated in order to avoid health and environmental concerns~\cite{RPJ17p424,MSEA420p171,AEM15p1014,JACS90p3720,IJACT7p803,MSF727p752}. 
The use of additives in the slurries affects directly its behavior as well as the properties of the tape cast substrates. 
A tape casting slurry must be adjusted in order to produce tapes with no defects, microstructural homogeneity and high mechanical strength after sintering. 
Thus, the composition and the rheological behavior of the aqueous slurries must be characterized and optimized in order to obtain green tapes, cracks and defects free with high and homogeneous green structure. 

Yttria-stabilized zirconia (YSZ) is intensively used as the electrolyte material in SOFC due its high ionic conductivity and chemical stability~\cite{MSF727p752,JPS195p2463,CI39p8279}. 
In particular, the YSZ is the most widely used electrolyte material for oxygen sensor and fuel cell applications. 
Pure zirconia ZrO$_{2}$ cannot act as a good electrolyte owing to its poor ionic conductivity and phase transformation (monoclinic/tetragonal) on heating associated with a large volume change. 
Doping of ZrO$_{2}$ with a small amount (3-10 mol$\%$) of a divalent or trivalent oxide can stabilized cubic fluorite phase and, in the process, increases its oxygen vacancy concentration leading to an enhanced ionic conductivity. 
This makes stabilized zirconia suitable for use as an electrolyte material and Yttria (Y$_{2}$O$_{3}$) is the most commonly used dopant for stabilizing zirconia for the aforementioned applications~\cite{CI27p731}.

At the same time, the functionalization of these tapes through insertion of metals and/or non-metallic alloys by using magnetron sputtering technique emerge as an attractive tool. 
The magnetron sputtering technique enables the growth of nanostructures in the single film or multilayer geometry for distinct technological applications. 
Traditionally, thin film are grown onto rigid amorphous or oriented substrates, and electrical and magnetic properties are explored for distinct purposes. 
In recent years, flexible substrates, with specific mechanical and electrical properties, have attracted considerable interest, mainly for flexible electronics devices~\cite{JMMM355p136,JMMM378p499,NRL7p230,MSEB211p115}. 
In this sense, flexible tapes with distinct compositions, such as ZrO$_{2}$, arise as a promising candidate for substrate. 
On the other hand, this same integration between tape casting and magnetron sputtering techniques can be used to functionalize the tape before sintering, changing the mechanical and electrical properties of the resulting ceramic materials. 
However, until this moment, a comparison of the structural and electrical and/or magnetic properties of nanostructures grown onto conventional and flexible tape substrates has not been reported yet, an achievement fundamental for future applications.

In this work, we associate tape casting and magnetron sputtering techniques to engineer flexible nanostructures and perform a systematic investigation of the structural, magnetic and electrical properties of NiFe/Cr/NiFe trilayer nanostructures, with variable thickness values of the NiFe and Cr layers, grown onto rigid glass and flexible ZrO$_2$ tape substrates.
We verify that structural properties, and magnetic and electric responses are reproduced, irrespective on the kind of employed substrate.
The results place the ZnO$_2$ tape as an attractive candidate for application as low-cost flexible substrate in the development of electrical and magnetic sensor elements with high sensitivity to mechanic stress.

\section{Experimental Procedure}
\label{Experiment}

\begin{figure}[!h]\centering
(a) \includegraphics[width=5.5cm]{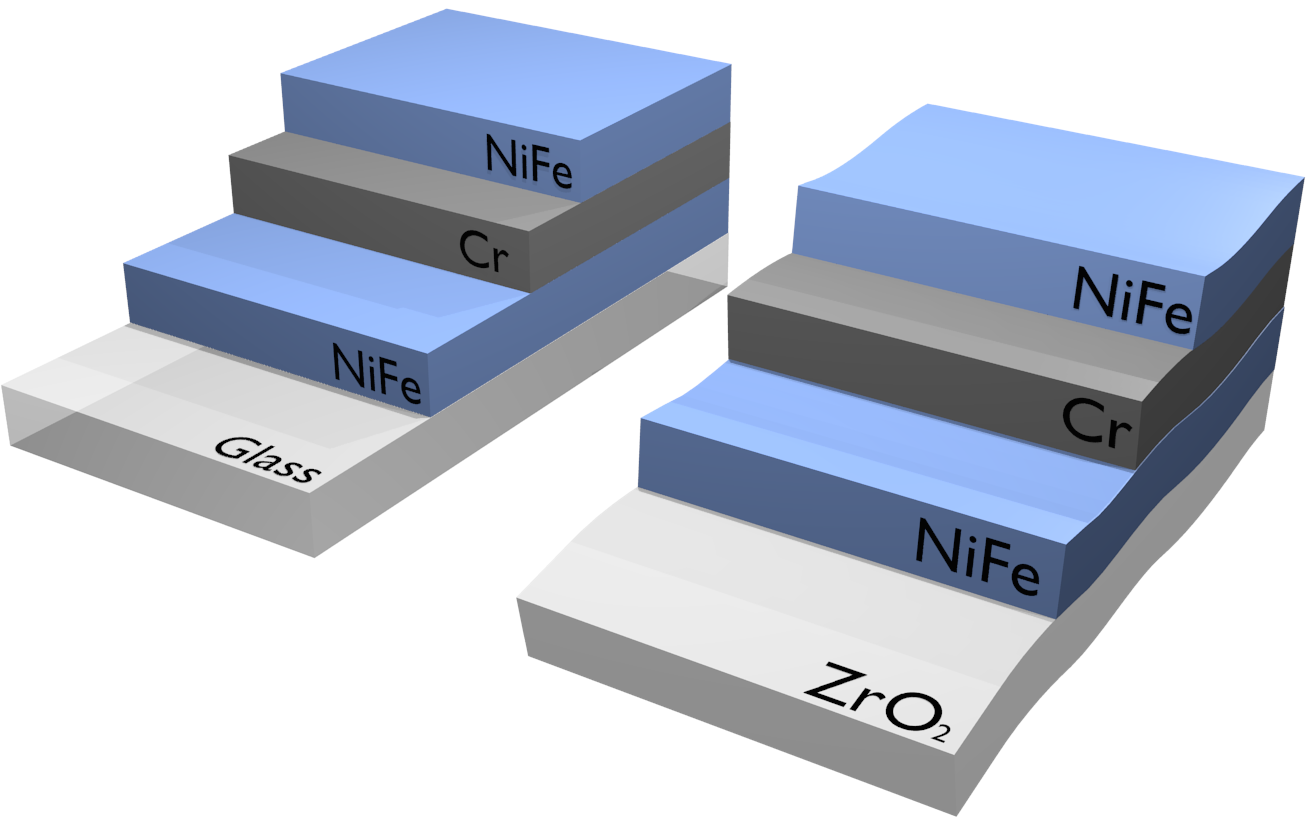} \vspace{0.3cm}  \\ 
 \includegraphics[width=5.5cm]{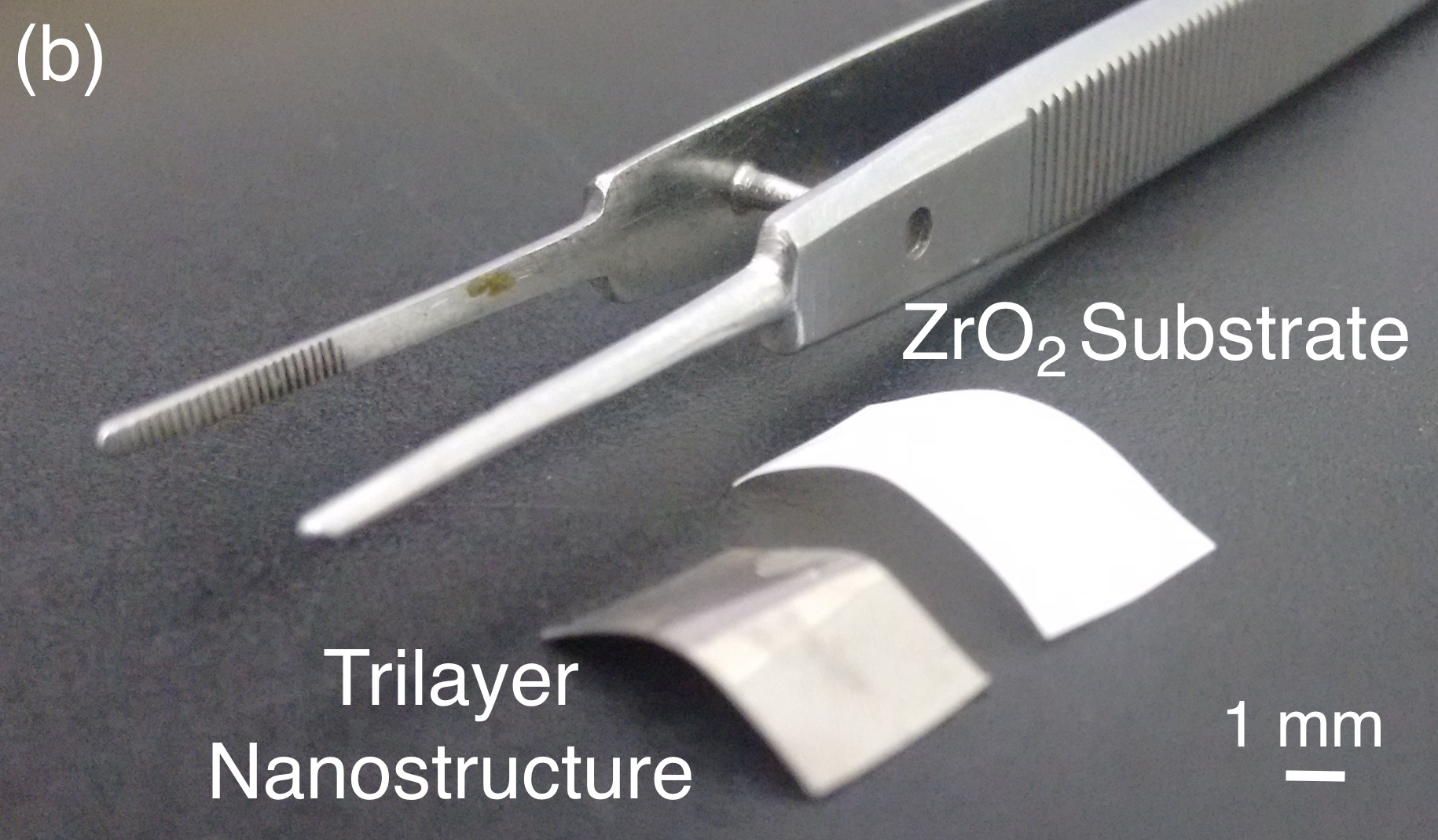} \vspace{-0.2cm} \hspace{0.45cm}
\caption {(a) Schematic representation of the NiFe/Cr/NiFe trilayer nanostructures grown onto rigid glass and flexible ZrO$_2$ tape substrates. (b) Flexible ZrO$_{2}$ tape substrate and a trilayer nanostructure grown onto this substrate. Notice the remarkable flexibility of the substrate.} \label{fig01}
\end{figure}

For the study, we consider a set of Ni$_{81}$Fe$_{19}$/Cr/Ni$_{81}$Fe$_{19}$ trilayer nanostructures, with variable thickness values of the NiFe and Cr layers, grown onto rigid glass and flexible ZrO$_2$ tape substrates. Figure~\ref{fig01} shows and schematic representation of the investigated samples, as well as the remarkable flexibility of the ZrO$_2$ tape substrate.

To obtain the flexible ZrO$_2$ tapes, slurries are prepared with $55$ wt.$\%$ YSZ powder ($3$YSZ), $3$ mol.$\%$ Y$_{2}$O$_{3}$-stabilized ZrO$_{2}$ (Tosoh Corporation, Japan) with average particle size of $400$~nm, $22$ wt.$\%$ deionized water used as solvent, and $1$ wt.$\%$ dispersant (Darvan 821A, Vanderbilt), and using alumina ball milling for $24$~h. Subsequently, $20$ wt.$\%$ binder (Monowilith LDM-6138, Clariant), $0.5$ wt.$\%$ antifoam (Antifoam A, Sigma-Aldrich) and $1.5$ wt.$\%$ surfactant (coconut diethanolamide, Stepan) are added to the mixture and the suspension is mixed for more $30$~min. The slurries are cast with a CC-$1200$ Mistler tape caster at $25^\circ$C and $5$~cm$\cdot$min$^{-1}$. The thickness of the tapes is around $125$~$\mu$m.

The stability of the yttria-stabilized zirconia slurries is analyzed by means of rheological characterization through viscosity curves. The viscosity experiments are carried out using a Haake Viscotester-Thermo Fischer Scientific viscosimeter with cone and plate geometry, at room temperature and with shear stress between $0$ and $1000$~s$^{-1}$.

With respect to the NiFe/Cr/NiFe trilayer nanostructures, the samples are produced by magnetron sputtering, with normal incidence, onto rigid glass and flexible ZrO$_2$ tape substrates with dimensions of $4\times 4$ mm$^{2}$, and
The films are deposited using the following parameters: base vacuum of $5.0 \times 10^{-6}$ Torr, deposition pressure of $3.0$ mTorr with a $99.99$\% pure Ar, and dc source with currents of $40$ mA and $60$ mA used for the Cr and NiFe layer deposition, respectively. Under these conditions, the deposition rates are $0,47$~\AA/s for Cr and $0.59$~\AA/s for NiFe alloy. 
Table~\ref{tab01} presents the complete set of trilayer nanostructures investigated in this work. Notice that to verify the magnetic and electrical properties with respect to thickness of NiFe layers, all the samples present total thickness of $[2 t_{\textrm{NiFe}}+ t_{\textrm{Cr}}] = 300$~nm, while the values of the thicknesses for the NiFe and Cr layers, $t_{\textrm{NiFe}}$ and $t_{\textrm{Cr}}$, respectively, varies continuously.
\begin{table}[!hb]
\vspace{-.0cm}
\centering 
\scriptsize
\caption{Thickness $t_{\textrm{NiFe}}$ and $t_{\textrm{Cr}}$ values of the NiFe and Cr layers composing the trilayer nanostructures grown onto rigid glass and flexible ZrO$_2$ tape substrates.}
\label{tab01}
\begin{tabular}{cccc} \hline\hline
Sample     & $t_{\textrm{NiFe}}$ (nm)       & $t_{\textrm{Cr}}$ (nm)         & $t_{\textrm{NiFe}}$ (nm)       \\  \hline 
T1   & 75.00   & 150.00     & 75.00   \\
T2   & 82.50    & 135.00      & 82.50 \\
T3   &  90.00    & 120.00      & 90.00 \\
T4   &  97.50    & 105.00      & 97.50 \\
T5   &  105.00    & 90.00      & 105.00 \\
T6   & 112.50   & 72.50     & 112.50  \\ 
T7   & 120.00   & 60.00     & 120.00                      \\ 
T8    & 127.50   & 45.00     & 127.50 \\
T9     & 135.00   & 30.00     & 135.00  \\
T10   & 142.50   & 15.00     & 142.50  \\ \hline \hline
\end{tabular}
\label{tab01}
\vspace{-0.0cm}
\end{table}

The structural properties of the trilayer nanostructures onto distinct substrates are verified by X-ray diffraction (XRD), which verify the structural character of the films and preferential growth direction. The measurements are obtained using a Rigaku Miniflex II diffractometer, in the Bragg-Brentano ($\theta-2\theta$) geometry, with Cu-K$_{\alpha}$ radiation. 

The magnetic behavior is investigated through magnetization curves measured at room temperature using a Lake Shore model $7404$ vibrating sample magnetometer, with maximum in-plane magnetic field of $\pm 300$~Oe. In particular, curves are acquired along the two lateral directions of the samples, to verify the magnetic anisotropy of the trilayer nanostructures. 

Finally, the electrical properties of the samples are studied through $I-V$ curves. In this case, the electrical response is obtained using a $238$ - Keithley High Current Source-Measure Unit through the measurement by four point probe method.

\section{Results and discussion}
\label{Results_and_discussion}

\begin{figure}[!h]\centering
\includegraphics[width=7.0cm]{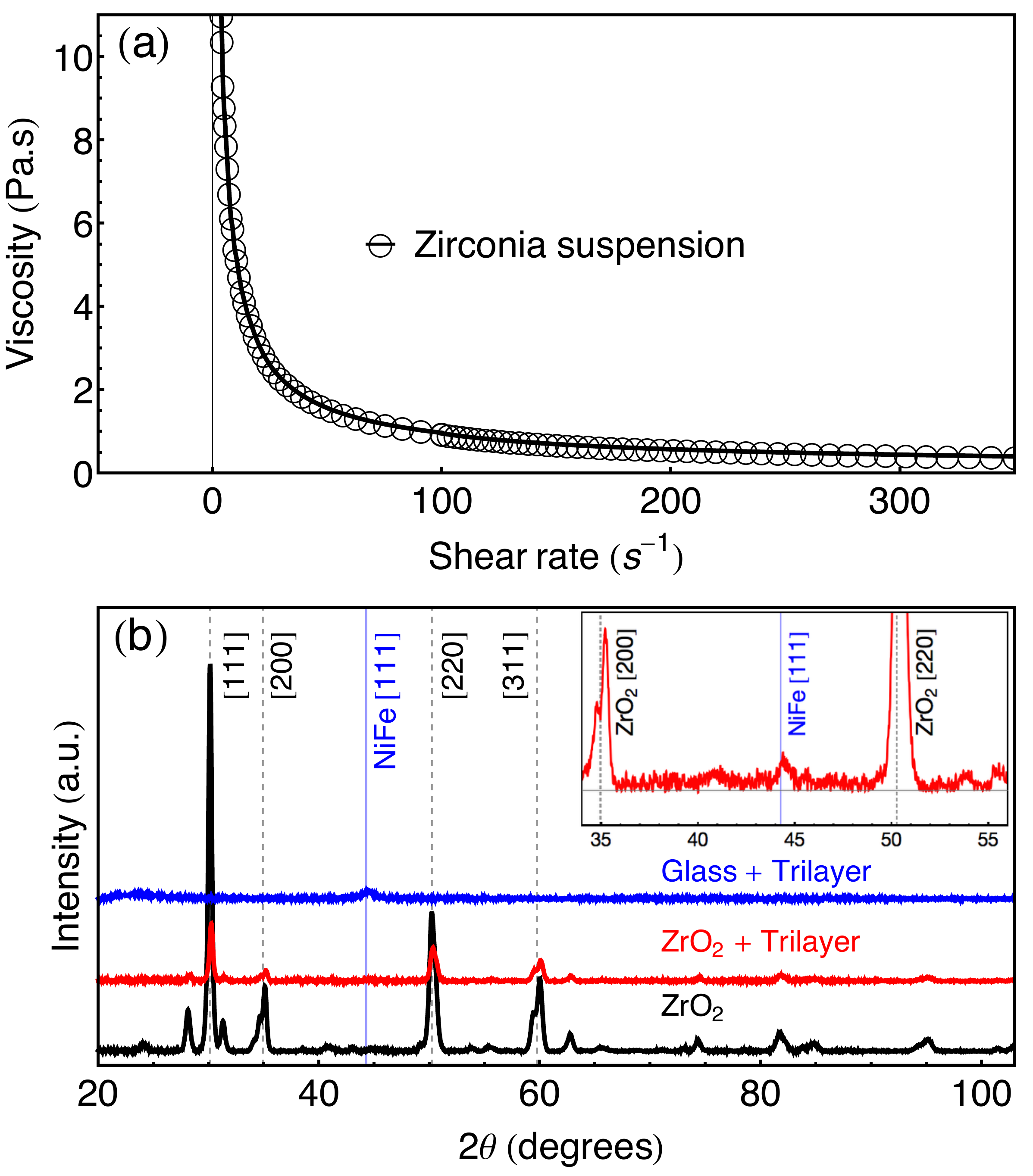} \vspace{-0.3cm}
\caption {(a) Viscosity as a function of the shear rate for the $3$YSZ suspension. 
(b) XRD results for the ZrO$_2$ tape and for the T5 trilayer nanostructures grown onto rigid glass and flexible ZrO$_2$ tape substrates. The standard patterns for ZrO$_{2}$ and Ni$_{81}$Fe$_{19}$ alloy are obtained respectively from the PDF cards No $00-500-0038$ (black dashed lines) and No $00-038-0419$ (blue solid lines). In the inset of (b), in detail, the NiFe $[111]$ peak in the XRD result for the trilayer nanostructure grown onto the flexible ZrO$_2$ tape substrate.}
\label{fig02}
\end{figure}

First of all, we perform the rheological characterization of the $3$YSZ suspension.
Figure~\ref{fig02}(a) shows the viscosity curve as a function of the shear rate. We observe a clear decrease of viscosity with the increasing shear rate, characterizing the pseudoplastic behavior of the suspension.
According to reports found in literature~\cite{Reed1995,Morrison2001,Dinger2002}, the pseudoplastic behavior is typical of suspensions presenting agglomerates. At low shear rates, interparticle attractive forces are predominant on hydrodynamic forces, leading to the formation of flakes. On the other hand, at high shear rates, hydrodynamic forces exerted by the flow become larger and, consequently, the flakes are destroyed, and water trapped inside them is gradually released to assist in the flow of particles, causing the reduction of the viscosity of suspension.
It is expected that suspensions for tape casting exhibit pseudoplastic behavior, enabling the production of homogeneous tapes with smooth surface, which contributes to the quality of the final product~\cite{CI38p2319}.

Figure~\ref{fig02}(b) shows the XRD result for the ZrO$_2$ tape, together with the ones obtained for selected trilayer nanostructures grown onto rigid glass and flexible ZrO$_2$ tape substrates. 
Regarding the structural properties, for the ZrO$_2$ tape, several peaks are identified and the pattern clearly indicates the polycrystalline state of the material.
The peak intensity associated to the ZrO$_2$ decreases for the trilayer nanostructure grown onto the tape, as expected. 
On the other hand, the pattern indicates the [111] texture for the NiFe alloy, assigned by peak identified at $2\theta = 44.28^\circ$ (See inset). Similar results are obtained for all trilayer nanostructures of the set, according to the substrate.
Thus, it is important to point out that the XRD patterns verified for trilayer nanostructures, irrespective on the substrate, reflect similar structural features for the NiFe, as frequently found for this alloy in the thin film geometry~\cite{APL104p102405}.  

\begin{figure}[!h]\centering
\includegraphics[width=8.0cm]{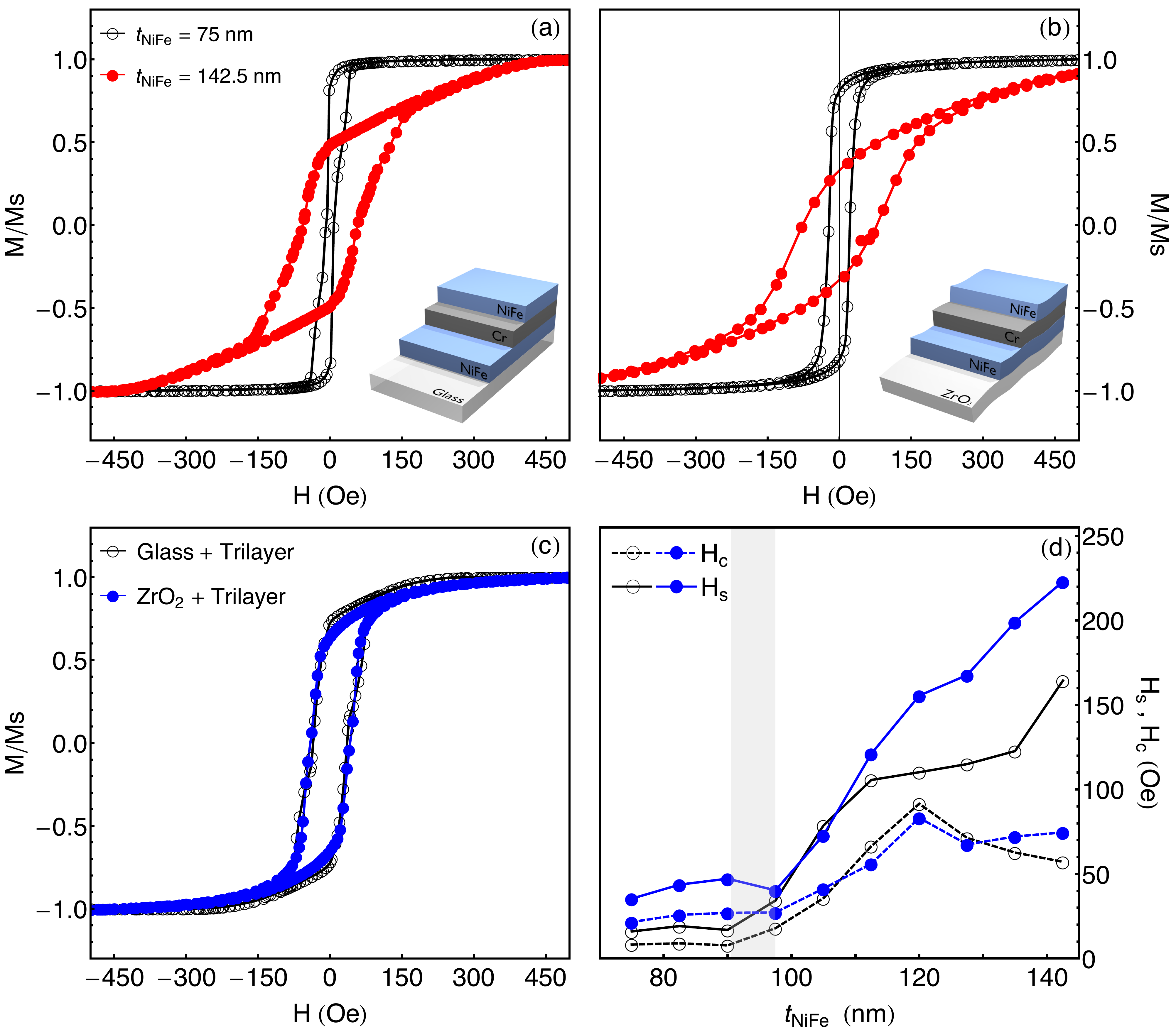}  \vspace{-0.5cm}
\caption {Normalized magnetization curves for the selected T1 and T10  trilayer nanostructures onto (a) rigid glass and (b) flexible ZrO$_2$ tape substrates. 
(c) Comparision of the magnetic behavior for the T5 samples grown onto distinct substrates. 
(d) Saturation field H$_{\textrm{s}}$ and coercive field H$_{\textrm{c}}$ as a function of thickness of a single NiFe layer of the nanostructure for the samples produced with distinct substrates. 
The gray zone represents the thickness range in which is verified the emergence of an out-of-plane magnetic anisotropy contribution.}
\label{fig03}
\end{figure}

With respect to the magnetic behavior, Fig.~\ref{fig03} present magnetization curves for selected trilayer nanostructure grown onto distinct substrates.
For all samples, the angular dependence of the curves (not shown here) indicate isotropy of the magnetic behavior in the plane. 
When analyzed as a function of the NiFe thickness, an evolution in the shape of the magnetization curves is noticed, indicating a thickness range which splits the samples in two groups according to the magnetic behavior, irrespective on the employed substrate.
The trilayer nanostructures with thickness smaller than $t_{\textrm{NiFe}} \approx 95$~nm present square magnetization loops (See Fig.~\ref{fig03}(a,b)), suggesting in-plane magnetization, without any out-of-plane component.
The domain wall motion is the main mechanism acting during the magnetization reversal, a fact revealed by the own shape of the curve, as well as by the almost constant high remanent magnetization and low saturation and coercive fields (Fig.~\ref{fig03}(d)).
For the trilayer nanostructures with thickness above $t_{\textrm{NiFe}}\approx 95$~nm, the curves are characterized by a magnetization reversal at fields close to the coercive field, occurring mainly through domain wall motion, followed by a linear approach to the magnetic saturation in which the main magnetization mechanism is the magnetization rotation. 
The drastic change of the shape of the magnetization curves reveals the deterioration of the soft magnetic properties and the appearance of an out-of-plane anisotropy contribution~\cite{PRB69p174402, JAP103p07e732,PB404p2784,JAP114p053908}, features directly related to the increase of the saturation and coercive fields as the NiFe thickness is raised. 
Despite the strong changes in the magnetic behavior with the NiFe thickness, notice the striking similarity between the curves obtained for trilayer nanostructures with distinct substrates, um fact clearly evidenced in the direct comparison of magnetization curves (Fig.~\ref{fig03}(d)), as well as in the behavior of $H_{\textrm{s}}$ and $H_{\textrm{c}}$ with NiFe thickness (Fig.~\ref{fig03}(c)). 

\begin{figure}[!h]\centering
\includegraphics[width=5.0cm]{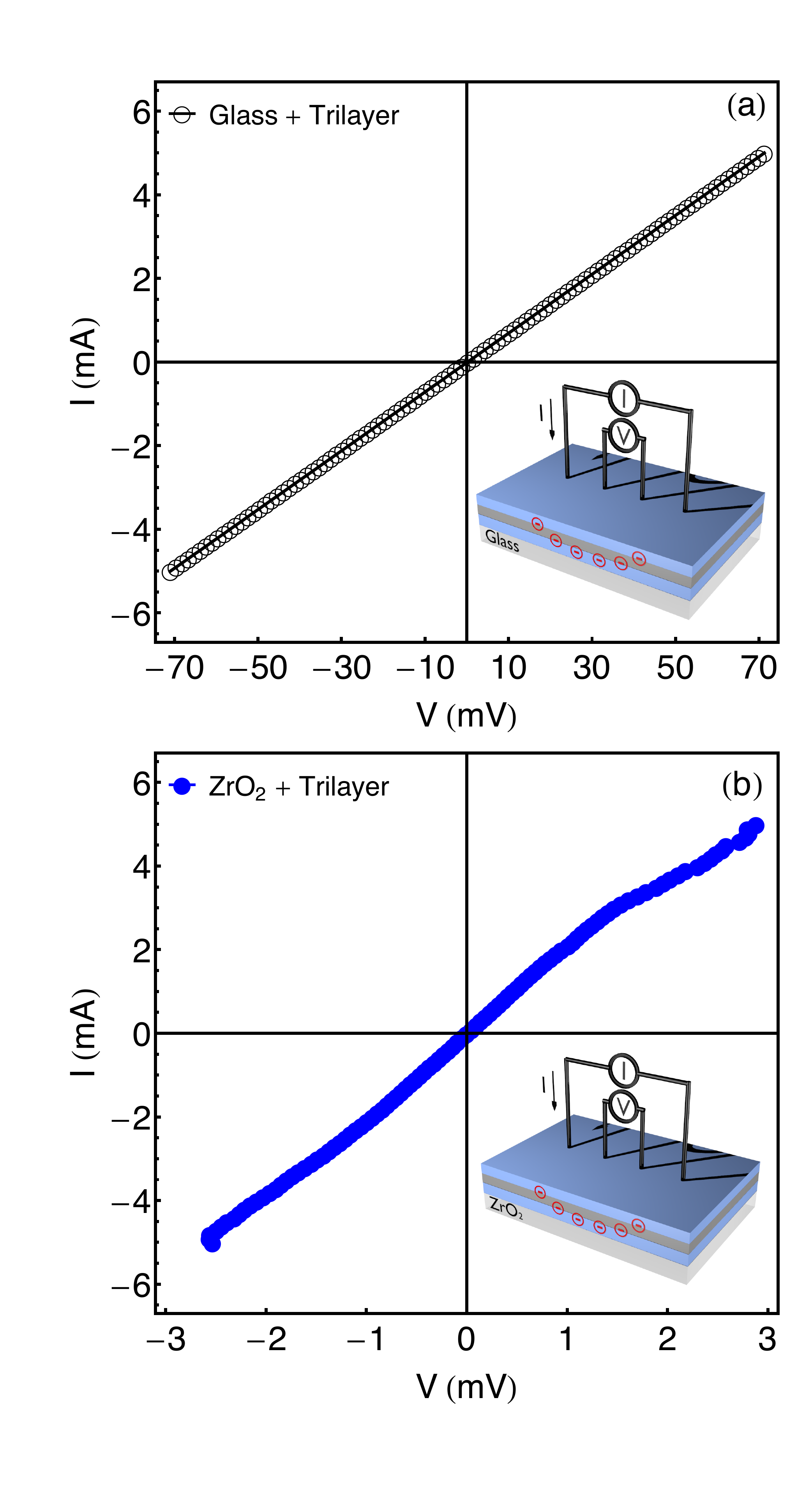} \vspace{-0.5cm}
\caption {$I-V$ curves for the selected T5 trilayer nanostructures onto (a) rigid glass and (b) flexible ZrO$_2$ tape substrates.}
\label{fig04}
\end{figure}

Finally, on the electrical response, Fig.~\ref{fig04} presents $I-V$ curves for selected trilayer nanostructures grown onto rigid glass and flexible ZrO$_2$ tape substrates. 
Although studies for distinct temperatures are not performed, it is possible to infer that the samples have ohmic behavior, since the electric resistance is constant, a consequence of the linear relation between current and voltage.
All the samples present similar behavior, although the resistance presents a clear dependence with the change of $t_{\textrm{NiFe}}$ and $t_{\textrm{Cr}}$, a consequence of the difference of resistivity of the materials.
Moreover, this behavior has a fundamental discrepancy for nanostructures grown onto glass and ZrO$_2$ tape, suggesting a dependence on the substrate. 
In this case, while for the trilayer nanostructures onto rigid substrate the resistance is very high, due to the electric resistivity of the insulating glass, the one onto the flexible tape present lower resistance, a signature of the smaller resistivity of the ZrO$_{2}$, as expected.

After all, our results raise important issues on the growth of trilayer nanostructures onto rigid glass and flexible ZrO$_2$ tape substrates. 
The structural properties, and magnetic and electric responses are reproduced, irrespective in the kind of employed substrate, corresponding to an important advance for the applicability of the ZrO$_2$ tape. 
Thus, the results place the ZrO$_2$ tape as an attractive candidate for application as low-cost flexible substrate in the development of electrical and magnetic sensor elements with high sensitivity to mechanic stress, in the case of a magnetostrictive alloy is employed.

\section{Conclusion}
\label{Conclusion}

In conclusion, we have shown the mirroring of the structural, magnetic and electrical properties of NiFe/Cr/NiFe trilayer nanostructures, with variable thickness values of the NiFe and Cr layers, grown onto rigid glass and flexible ZrO$_2$ tape substrates. 
The fact that the trilayer nanostructures can be reproduced in distinct kinds of substrates corresponds to an important advance for their applicability. 
The results place the ZrO$_2$ tape as an attractive candidate for application as low-cost flexible substrate in the development of electrical and magnetic sensor elements with high sensitivity to mechanic stress.
Moreover, from now, a broad range of possibilities for applications can be explored. 
In particular, the mirroring of the main features makes the association of tape casting and magnetron sputtering techniques an interesting route to insert specific elements in the tape before sintering, improving/modifying the resulting ceramic material according to the applicability, such as in solid oxide fuel cells. 

\section*{Acknowledgments}
The research is supported by the Brazilian agencies CNPq, CAPES, and FAPERN.

\bibliographystyle{model1a-num-names}
\bibliography{References_MI.bib}

\end{document}